\documentstyle[12pt]{article}
\begin{document}

\title{1/N expansion formalism for high-spin states}

\author{S. Kuyucak$^1$ and S.C. Li \\
Department of Theoretical Physics,\\ Research School of Physical Sciences,\\
Australian National University,
Canberra, ACT 0200, Australia\\}
\date{}

\maketitle

\begin{abstract}
The 1/$N$ expansion solutions for the interacting boson model are extended
to higher orders using computer algebra. The analytic results are
compared with those obtained from an exact diagonalization of the Hamiltonian
and are shown to be very accurate.
The extended formulas  for level energies and E2 transitions will be
useful in the analysis of high-spin states in both normal and superdeformed
nuclei.
\end{abstract}

\noindent
$^1$ E-mail: sek105@phys.anu.edu.au
\vfill \eject

\baselineskip=21pt

Application of the interacting boson model \cite{iac87} (IBM) to the
high-spin states of normal and superdeformed nuclei has been rather slow, and
certainly not commensurate with the current experimental activity in this
field.
The reasons for this are i) the necessity of including $g$
bosons in the basis in addition to the usual $s$ and $d$ bosons,
and ii) the need to include broken pairs (quasi-particles) in the basis
if backbending occurs.
In this work, we address the former problem and refer to Ref. \cite{cas94}
for recent developments in the latter.
The importance of $g$ bosons has been well established in studies
involving low-lying levels of deformed nuclei (see Refs. \cite{dev92,hey93}
for reviews).
There are also many experiments on high-spin states
where results are compared to the $sd$-IBM calculations showing its
inadequacy.
The difficulty with performing $sdg$-IBM calculations for deformed nuclei
(with boson numbers $N=12-16$)
is that the basis space is too large, and numerical diagonalization
of a Hamiltonian is not possible even on a supercomputer.
Therefore various truncation schemes have been devised.
These schemes are geared towards reproducing the
low-lying spectrum and can not be expected to give reliable results for the
high-spin states.
The situation gets even worse in the case of superdeformed nuclei where
recent microscopic studies have shown that the number of bosons
is around $N=30-40$ \cite{hon92,hon93},
which renders any attempt to numerical diagonalization futile.

A more promising approach is to use the angular momentu"m projected mean field
theory which leads to a 1/$N$ expansion for all matrix elements (m.e.) of
interest \cite{kuy88}. Initial calculations in the 1/$N$ expansion formalism
were restricted to order 1/$N^2$ which was sufficient for low-spin ($L<N$)
phenomenology. For an accurate description of high-spin properties
(e.g. dynamic moments of inertia), however, one needs to include terms up to
order 1/$N^6$ (note that these terms actually go as $(L/N)^6$, and are not
small for high-spin states).
Hand calculation of m.e. to such high orders is impractical
and has never been attempted. Recent advances in computer algebra have
finally broken this impasse.
The purpose of this letter is to report on the results of such a 1/$N$
expansion
calculation facilitated by the Mathematica software \cite{wol91}.
Applications are made to the superdeformed bands in Hg isotopes.

We consider a general formulation of the IBM and introduce the boson creation
and annihilation operators $b^\dagger_{l\mu}, b_{l\mu}$ with $l=0,2,4, \ldots$.
The ground band can be written as a condensate of intrinsic bosons as
\begin{equation}
|N,{\bf x}\rangle =(N!)^{-1/2}(b^\dagger)^N|0\rangle,\quad
b^\dagger=\sum_{l} x_{l} b^\dagger_{l0},
\end{equation}
where $x_{l}$ are the normalized boson mean fields which are associated
with the deformation parameters of the system. Here we assume that the system
is axially symmetric, which is a good approximation as will be seen later.
The other bands can be obtained from the ground band by acting with the
orthogonal intrinsic boson operators.
A general boson Hamiltonian with one- and two-body terms is given by
\begin{eqnarray}
&&H = \sum_l \varepsilon_l \hat n_l + \sum_{k=0}^{2l_{\max}}
\kappa_k T^{(k)}\cdot T^{(k)}, \nonumber \\
&&\hat n_l = \sum_\mu b^\dagger_{l\mu} b_{l\mu}, \quad
T^{(k)} = \sum_{jl} t_{kjl}[b_j^\dagger \tilde b_l]^{(k)},
\label{ham}
\end{eqnarray}
where brackets denote tensor coupling of the boson operators,
$\tilde b_{l\mu}=(-1)^{\mu}b_{l-\mu}$, and
$\hat n_l$ and $T^{(k)}$ are the boson number and multipole operators,
respectively. The parameters in the model are the single boson energies
$\varepsilon_l$, the multipole strengths $\kappa_k$, and the coefficients
$t_{kjl}$. For consistency, the same multipole operators are used in the
calculation of electromagnetic transition rates.

Matrix elements of any operator with angular momentum projection can be
evaluated efficiently using boson calculus and angular momentum algebra
techniques \cite{kuy88}. We refer to the original references for details
and give here the intermediate results that form the basis of the Mathematica
calculation. The exact expectation values of the number and multipole
operators (\ref{ham}) in the ground band are given by
\begin{eqnarray}
&&\hskip -.5cm \langle\hat n_l\rangle_L = {N x_l^2 \over F(N,L)}
\sum_I \langle L0l00|I0\rangle^2 F(N-1,I), \nonumber\\
&&\hskip -.5cm \langle T^{(k)}\cdot T^{(k)}\rangle_L = {2k+1\over F(N,L)}
\biggl\{ N \sum_{jl} {(t_{kjl} x_l)^2 \over 2l+1}
\sum_I \langle L0l00|I0\rangle^2 F(N-1,I) \nonumber\\
&&\hskip 4.cm + N(N-1) \sum_{jlj'l'J} t_{kjl} t_{kj'l'} x_j x_l x_{j'} x_{l'}
\langle j0j'0|J0\rangle \langle l0l'0|J0\rangle \nonumber\\
&&\hskip 4.cm \times
\left\{ \begin{array}{ccc} j & j' & J\\l' & l & k\end{array}\right\}
\sum_I \langle L0J0|I0\rangle^2 F(N-2,I) \biggr\}.
\label{me}
\end{eqnarray}
Here $F(N,L)$ denotes the normalization integral which has the 1/$N$ expansion
\cite{kuy94}
\begin{equation}
F(N,L)={1\over N} \sum_{n=0} {(-1)^n\over n!(aN)^n}
\sum_{m=0}^n \alpha_{nm} \bar L^m,
\label{norm}
\end{equation}
where bar denotes the angular momentum eigenvalues, $\bar L\equiv L(L+1)$ and
$a=\sum_l \bar l x_l^2$ represents the ``average
angular momentum squared" carried by a single boson.
The coefficients $\alpha_{nm}$ in Eq. (\ref{norm}) have been calculated
recently
utilizing the Mathematica software \cite{kuy94}.
It is clear from the expressions (\ref{me},\ref{norm}) that, in principle, one
can evaluate these m.e. to any desired order in 1/$N$ (see Ref. \cite{kuy88}).
However, the amount of algebraic manipulation required grows
exponentially with each order, and one is soon forced to give up.
In contrast, Mathematica is very efficient at such ``algebra crunching".

Before presenting the final results, it will be useful to comment on the
general
form of the m.e. of a $k$-body operator $\hat O$, and illustrate
the concept of layers in the 1/$N$ expansion
(see Ref. \cite{kuy94} for more details)
\begin{equation}
\langle\hat O\rangle_L = N^k \sum_{n,m} {O_{nm}\over (aN)^m}
\Bigl({\bar L \over a^2N^2}\Bigr)^n.
\label{me1}
\end{equation}
The expansion coefficients $O_{nm}$ in Eq. (\ref{me1}) involve various
quadratic
forms of the mean fields $x_{l}$ corresponding to the single-boson m.e. of
$\hat O$ and its moments. The terms with $n+m-1=i$, constant, are referred as
the $i$th layer. An expansion in layers rather than in 1/$N$ is preferred
on both practical and physical grounds. This is because terms in the same
layer have similar forms (although they have different $N$ dependence),
while complexity of terms grows exponentially with increasing layers.
For example, a complete calculation to order 1/$N^6$ would involve terms
belonging to the fourth, fifth and sixth layers which are very complicated
yet completely unnecessary as their contribution to the m.e. is beyond the
desired level of accuracy. Here we restrict ourselves to a third layer
calculation of the m.e. which is found to be sufficiently accurate for
description of high-spin states.
For the one-body m.e. in Eq. (\ref{me}), one obtains
\begin{eqnarray}
&&\hskip -1cm \langle\hat n_l\rangle_L= Nx_l^2 \Bigl\{
1 + {1 \over aN}\Bigl(a-\bar l\Bigr)
+ {1 \over (aN)^2}\Bigl(-a + a_1/2 + (1 - a_1/a)\bar l + \bar l^2/2\Bigr)
\nonumber\\
&&\hskip 1cm + {1 \over (aN)^3}\Bigl(
a + 2a^2 - 7a_1/3 - aa_1 + 5a_1^2/4a - a_2/3 \nonumber\\
&&\hskip 2.2cm +  (-1 - 2a + 2a_1+ 7a_1/2a- 5a_1^2/2a^2 + a_2/2a)\bar l
\nonumber\\
&&\hskip 2.2cm + (-7/6 - a + 5a_1/4a)\bar l^2 - \bar l^3/6 \Bigr) \nonumber\\
&&+ {\bar L \over (aN)^2}\Bigl[ (-a+\bar l)
+ {1 \over aN}\Bigl(2a+2a^2-2a_1+(-2-2a+3a_1/a)\bar l-\bar l^2 \Bigr)
\nonumber\\
&&\hskip 1cm + {1 \over (aN)^2}\Bigl(-3a-12a^2-4a^3+21a_1/2+11aa_1
-15a_1^2/2a+3a_2/2  \nonumber \\
&&\hskip 2.2cm  + (3+12a+4a^2-33a_1/2-14a_1/a+25a_1^2/2a^2-2a_2/a)\bar l
\nonumber \\
&&\hskip 2.2cm  + (7/2+11a/2-5a_1/a)\bar l^2 + \bar l^3/2 \Bigr) \Bigr]
\nonumber \\
&&+ {\bar L^2\over 2(aN)^4}\Bigl[
(-a-2a^2+3a_1/2+(1+2a-2a_1/a)\bar l+\bar l^2/2) \nonumber\\
&&\hskip 1cm + {1 \over aN}\Bigl( 4a+21a^2+14a^3-16a_1-51aa_1/2+13a_1^2/a-2a_2
\nonumber\\
&&\hskip 2.2cm +   (-4-21a-14a^2+34a_1+20a_1/a-39a_1^2/2a^2+5a_2/2a)\bar l
\nonumber\\
&&\hskip 2.2cm + (-4-17a/2+13a_1/2a)\bar l^2-\bar l^3/2 \Bigr) \Bigr]
\nonumber\\
&&+ {\bar L^3\over 3(aN)^6}\Bigl[
-a-6a^2-6a^3+25a_1/6+9aa_1-15a_1^2/4a+5a_2/12 \nonumber\\
&&\hskip 2.2cm +  (1+6a+6a^2-45a_1/4-5a_1/a+21a_1^2/4a^2-a_2/2a)\bar l
\nonumber\\
&&\hskip 2.2cm + (5/6+9a/4-3a_1/2a)\bar l^2+\bar l^3/12 \Bigr] \Bigr\},
\label{1body}
\end{eqnarray}
where $a_n=\sum_l \bar l^{n+1} x_l^2$ denotes the higher moments of $a$.
We have checked Eq. (\ref{1body}) against two results: i) it satisfies the
number conservation, i.e. $\sum_l \langle \hat n_l \rangle=N$,
ii) it reproduces the analytic formulas available in the SU(3) limit
\cite{iac87}. A similar Mathematica evaluation of the two-body m.e. in Eq.
(\ref{me}) gives \begin{eqnarray}
&&\hskip -1cm \langle T^{(k)}\cdot T^{(k)}\rangle_L=
N^2 \Bigl\{ U_k + {1 \over aN}\Bigl(aU_k - U_k1 + aC_k\Bigr) \nonumber\\
&&\hskip 2.3cm + {1 \over (aN)^2}\Bigl((-2a+a_1)U_k + (1-a-a_1/a)U_{k1} +
    U_{k2}/2 + a^2C_k-aC_{k1}\Bigr) \nonumber\\
&&\hskip 2.3cm + {1 \over (aN)^3}\Bigl((2a + 2a^2 - 14a_1/3 - aa_1+5a_1^2/2a-
2a_2/3)U_k\nonumber\\
&&\hskip 3.7cm + (-1 + a - a_1/2 + 7a_1/2a - 5a_1^2/2a^2+a_2/2a)U_{k1}
\nonumber\\
&&\hskip 3.7cm + (-7/6 + 5a_1/4a)U_{k2} - U_{k3}/6 \nonumber\\
&&\hskip 3.7cm + (-a^2 + aa_1/2)C_k + (a - a_1)C_{k1} + aC_{k2}/2\Bigr)
\nonumber\\
&&\quad + {\bar L \over (aN)^2}\Bigl[ -2aU_k + U_{k1} \nonumber\\
&&\hskip 1.7cm +  {1 \over aN}\Bigl((4a + 2a^2 - 4a_1)U_k
+ (-2 + a + 3a_1/a)U_{k1} - U_{k2} - a^2C_k + aC_{k1}\Bigr)  \nonumber\\
&&\hskip 1.7cm+ {1 \over (aN)^2}\Bigl((-6a - 16a^2 - 4a^3 + 21a_1 + 15aa_1 -
      15a_1^2/a + 3a_2)U_k \nonumber\\
&&\hskip 3.2cm + (3+2a-2a^2-4a_1-14a_1/a + 25a_1^2/2a^2 - 2a_2/a)U_{k1}
\nonumber\\
&&\hskip 3.2cm + (7/2+2a-5a_1/a)U_{k2}+U_{k3}/2 \nonumber\\
&&\hskip 3.2cm + (2a^2 + 2a^3-2aa_1)C_k+(-2a-2a^2+3a_1)C_{k1} -aC_{k2}\Bigr)
\Bigr] \nonumber\\
&&\quad  + {\bar L^2\over 2(aN)^4}\Bigl[ (-2a - 2a^2 + 3a_1)U_k +
(1 - 2a_1/a)U_{k1} + U_{k2}/2 \nonumber\\
&&\hskip 2.cm+ {1 \over aN}\Bigl((8a + 30a^2 + 14a^3 - 32a_1 - 37aa_1 +
               26a_1^2/a - 4a_2)U_k \nonumber\\
&&\hskip 3.2cm + (-4 - 8a + 2a^2 + 29a_1/2 + 20a_1/a - 39a_1^2/2a^2 +
                  5a_2/2a)U_{k1} \nonumber\\
&&\hskip 3.2cm + (-4-9a/2 + 13a_1/2a)U_{k2} - U_{k3}/2 \nonumber\\
&&\hskip 3.2cm + (-a^2 - 2a^3 + 3aa_1/2)C_k + (a + 2a^2 - 2a_1)C_{k1}
+ aC_{k2}/2\Bigr) \Bigr] \nonumber\\
&&\quad + {\bar L^3\over 3(aN)^6}\Bigl[ (-8a - 36a^2 - 24a^3 + 100a_1/3
+ 54aa_1 - 30a_1^2/a + 10a_2/3)U_k \nonumber\\
&&\hskip 3.2cm + (4 + 12a - 24a_1 - 20a_1/a + 21a_1^2/a^2 - 2a_2/a)U_{k1}
\nonumber\\
&&\hskip 3.2cm + (10/3 + 6a - 6a_1/a)U_{k2} + U_{k3}/3\Bigr] \Bigr\}.
\label{2body}
\end{eqnarray}
Here the quadratic forms $C_{kn}$ arise from normal ordering and simulate
an effective one-body term
\begin{equation}
\quad C_{kn} = (2k+1) \sum_{jl} \bar{l}^n (t_{kjl} x_l)^2/(2l+1),
\end{equation}
while $U_{kn}$ represent the genuine two-boson interaction
\begin{equation}
U_{kn}=\sum_{jlj'l'I} \bar I^n \langle j0j'0|I0\rangle \langle l0l'0|I0\rangle
\left\{ \begin{array}{ccc} j & j' & I\\l' & l & k\end{array}\right\}
t_{kjl} t_{kj'l'} x_j x_l x_{j'} x_{l'}.
\end{equation}
For a given multipole, these sums can be evaluated in closed form using
Mathematica. For the quadrupole interaction, the first four terms needed in
Eq. (\ref{2body}) are given by
\begin{eqnarray}
&&U_2  = A^2,\quad U_{21} = (2A_1-3A)A,\nonumber\\
&&U_{22} = (2A_2-24A_1+18A)A + (A_{11}-A_2+7A_1)A_1 + (A_{11}-A_2)^2/12,
\nonumber\\
&&U_{23} = (2A_3-36A_2-18A_{11}+192A_1-144A)A \nonumber\\
&&\hskip 1cm +(3A_{21}-3A_3+48A_2+24A_{11}-146A_1)A_1/2\nonumber\\
&&\hskip 1cm -(3A_{21}-3A_3+25A_2-14A_{11})A_2/12 \nonumber\\
&&\hskip 1cm +(3A_{21}-3A_3+11A_{11})A_{11}/12, \label{un}
\end{eqnarray}
where the quadratic forms $A_{mn}$ in (\ref{un}) are defined as
\begin{equation}
A_{mn}=\sum_{jl} \bar{j}^m \bar{l}^n \langle j0l0|20\rangle t_{2jl} x_j x_l,
\label{amn}
\end{equation}
and correspond to various moments of the single-boson quadrupole m.e.
(note that the zero subscripts are suppressed for convenience).
The quadrupole m.e. given by (\ref{2body}-\ref{un}) reproduces the well known
Casimir eigenvalues in the SU(3) limit, hence also passes the SU(3) test.

To obtain the final energies, one has to perform variation after projection
(VAP) and determine the mean fields $x_l$ for each $L$. This can be done
algebraically using an ansatz for $x_l$ similar to Eq. (\ref{me1}), which
leads to sets of linear equations for the higher order terms \cite{kuy88}.
As the algebraic solution of the VAP equations are rather lengthy, we defer
their discussion to a longer paper.
Alternatively, one can determine $x_l$ numerically using, for example,
the simplex method. Both methods lead to equally accurate results.
Expressions for the excited band energies are obtained in a similar manner,
though they are somewhat more complicated due to contributions from
orthogonality and band mixing effects, and will be given in a future paper.

The other experimentally relevant quantities in the study of high-spin states
are the $E2$ transitions. In this case, the first layer
results are found to be sufficiently accurate, hence higher order calculations
are not necessary. For completeness, we quote the $E2$ transition m.e. in the
ground band \cite{kuy90}
\begin{eqnarray}
&& \langle L' \parallel T^{(2)} \parallel L \rangle =
N \hat L \langle L0\, 20|L'0\rangle
\Bigl\{ A - {1 \over aN} \bigl(A_1-3A-aA\bigr) \nonumber\\
&&\hskip 3cm -{1 \over 8(aN)^2} \Bigl[(A_2-A_{11}-10A_1+12A+4aA)(\bar L'
+\bar L) \nonumber\\
&& \hskip 4.5cm -{1\over 6} (A_2-A_{11}-6A_1+6A)(\bar L'-\bar L)^2\Bigr]
\Bigr\},
\label{e2}
\end{eqnarray}
where $\hat L = [2L+1]^{1/2}$ and $A_{mn}$
are defined in (\ref{amn}). Further expressions for in- and inter-band
transitions among the ground, $\beta$ and $\gamma$ bands, including band mixing
effects, are given in Ref. \cite{kuy90}. A fortran code for the 1/$N$
expansion calculation of energies and $E2$ transitions is available from the
authors upon request \cite{kl94}.

Before applying the 1/$N$ expansion results, we compare them with those
obtained from an exact diagonalization of the Hamiltonian \cite{san94}.
For this purpose we choose two different parametrizations of $sdg$-IBM:
(a) a pure quadrupole interaction which is appropriate for superdeformed
states \cite{hon92,hon93},
(b) quadrupole interaction plus a $g$ boson energy which reflects weaker
coupling of $g$ bosons in normal deformed states.
In both cases, the quadrupole parameters
$q_{22},q_{24}, q_{44}$ are scaled from their SU(3) values with a single
factor $q$ ($q_{02}=1$, $q_{jl}=t_{2jl}$). The $q$ factor gives a simple
measure for the breaking of the SU(3) symmetry which is necessary for a
realistic description of deformed nuclei.
Of necessity, the boson number is fixed at $N=10$, which requires some scaling
of the parameters from their realistic values at larger $N$.
Fig. 1 presents comparisons for the ground band energies.
The top figures show $E_L/\bar L$ as a function of $\bar L=L(L+1)$. In case
(a),
the agreement between the third layer 1/N expansion calculation (solid line)
and the exact results is excellent up to the maximum spin $L=4N$.
The second layer results (dotted line) start deviating from the exact ones
around $L=2N$ as do the third layer results with VBP (dashed line).
In case (b), the agreement
is still very good, with only a few percent difference at the very high-spin
region. Here, the VBP calculation shows even larger deviations,
underscoring the importance of the VAP procedure for high-spin states.
In the bottom figures, we compare the dynamic moment of inertia
${\cal J}^{(2)}$ which is much more sensitive to changes in structure.
Thus, the inadequacy of the second layer calculations, which is not very clear
in the energy plots, becomes obvious in ${\cal J}^{(2)}$ plots. In Fig. 2, we
compare the first layer results for the $E2$ transition m.e. (\ref{e2}) with
the exact ones.
The agreement is again very good up to very high-spins in both cases.
The test cases discussed above indicate that the extended formalism can be
applied with confidence in the spin region $L=N$-$3N$ which covers the
presently
available high-spin data. This, incidentally, also shows
the adequacy of the axial symmetry assumption made in the beginning.

The analytic expressions derived for energies and $E2$ transitions will be
useful in the study of high-spin states in both normal and superdeformed
nuclei. Here, we present an application of the 1/$N$ expansion formalism to
superdeformation which is more topical and harder to treat by numerical
diagonalization.
In super IBM, as proposed by Otsuka and Honma \cite{hon92,hon93}, normal bosons
are supplemented with superdeformed bosons which correspond to the Cooper-pairs
in the superdeformed potential. The number of superdeformed bosons, $N_{super}$
is typically around 30-40, and because of large deformation, g boson effects
are important. Thus, the super IBM offers a fertile ground for the application
of the 1/$N$ expansion.
We use the energy formula (\ref{2body}) to describe the superdeformed bands in
the Hg isotopes. The dynamic moments of inertia, ${\cal J}^{(2)}$ that result
from the quadrupole Hamiltonian are shown in Fig.~3. The three quadrupole
parameters $q_{22},q_{24}, q_{44}$ are scaled from their SU(3) values with a
single factor $q$. $N_{super}$ is determined from microscopic calculations,
\cite{hon92,hon93} and $\kappa$ and $q$ are fitted to the experimental data. A
good description of experimental ${\cal J}^{(2)}$ (circles) is obtained. We
note that the SU(3) limit corresponds to a rigid rotor and would give a flat
line for ${\cal J}^{(2)}$. This happens because in the SU(3) limit, the mean
fields $x_l$ are constant (independent of $L$), and the structure does not
change with rotation. In reality, one expects a gradual change in ${\cal
J}^{(2)}$ due to loss of pairing. The above study shows that this can be
simulated by breaking of the SU(3) symmetry which results in migration of the
mean fields from $s$ to $d$, and to $g$ bosons with increasing spin. The $q$
values obtained in the above fits indicate that this breaking is around 30\%.
It has been suggested that the identical band phenomenon may be due to an
underlying symmetry \cite{iac91,jan92}. It would be of interest to pursue this
suggestion by extending the present calculations to other bands and also to odd
nuclei.

Another area where the 1/$N$ expansion formalism could contribute
significantly is the study of high-spin states in normal deformed nuclei.
In many experiments on high-spin states, results were compared to the $sd$-IBM
calculations with negative connotations. This is presumably due to lack of
$sdg$-IBM calculations which, hopefully, will become more accessible
with the analytic formulas presented here.

This research was supported by the Australian Research Council, and in part
by an AAS/JSPS exchange grant. Numerical calculations were performed using the
Fujitsu VP of the ANU Supercomputer Facility.
S.K. thanks T. Otsuka and M. Honma for valuable discussions on super IBM,
and the members of the nuclear theory group at the University of Tokyo for
their hospitality.

\eject

\vfill \eject
{\Large \bf Figure captions}
\\[.5cm]
Fig. 1. Comparison of the ground band energies (top) and dynamic
moments of inertia (bottom) obtained from the
1/$N$ expansion with the exact numerical results (circles).
The different lines refer to the third layer calculation with VAP (solid line),
third layer with VBP (dashed line), and second layer with VAP (dotted line).
The VBP results for ${\cal J}^{(2)}$ deviate strongly from the exact ones
and are not shown to avoid cluttering of the figures.
The parameters of $H$ are $\kappa = -20$ keV, $q = 0.7$ for the
quadrupole interaction in both (a) and (b), and $\varepsilon_g=500$ keV in (b).
\\[.4cm]
Fig. 2. Comparison of the $E2$ transition m.e., Eq. (12) with the exact
diagonalization results (circles).
Solid line refers to VAP and dashed line to VBP.
\\[.4cm]
Fig. 3. Comparison of the experimental dynamic moment of inertia in
$^{190-194}$Hg (circles) with the super IBM calculations (solid lines).
The data are from \cite{sddata}. The parameters used in the fits
are $N_{super} = 29, 30, 31$, $\kappa = -35, -34, -33$ keV,
$q = 0.68, 0.72, 0.72$ for $^{190-192-194}$Hg, respectively.


\begin{thebibliography}{99}

\bibitem{iac87} F. Iachello and A. Arima, The interacting
boson model (Cambridge University Press, Cambridge, 1987).
\bibitem{cas94} R.F. Casten et al. eds., Proc. int. conf. on perspectives for
the interacting boson model (Padova, June 1994), (World Scientific,
Singapore, 1994).
\bibitem{dev92} Y.D. Devi and V.K.B. Kota, Pramana-J. Phys. 39 (1992) 413.
\bibitem{hey93} K. Heyde, in: Algebraic approaches to nuclear structure,
ed. R.F. Casten (Hardwood Academic Publishers, Switzerland, 1993) p. 323.
\bibitem{hon92} T. Otsuka and M. Honma, Phys. Lett. B 268 (1992) 305.
\bibitem{hon93} M. Honma, Superdeformation in the Hg-Pb region and an
extension of IBM, Ph.D. thesis, University of Tokyo, 1993;
M. Honma and T. Otsuka, in Ref. [2], p. 343.
\bibitem{kuy88} S. Kuyucak and I. Morrison, Ann. Phys. (N.Y.) 181 (1988) 79;
ibid, 195 (1989) 126.
\bibitem{wol91} S. Wolfram, Mathematica (Addison-Wesley, Redwood City, 1991).
\bibitem{kuy94} S. Kuyucak and K. Unnikrishnan, J. Phys. A, in press;
S. Kuyucak, in Ref. [2], p. 143.
\bibitem{kuy90} S. Kuyucak and I. Morrison, Phys. Rev. C41 (1990) 1803.
\bibitem{kl94} S. Kuyucak and S.C. Li, Computer code IBM-1/N (ANU, 1994).
\bibitem{san94} I. Morrison, Computer code SDGBOSON (University of Melbourne,
1986); SDGBOSON for the supercomputer, S.C. Li (ANU, 1993).
\bibitem{sddata} J.E. Draper et al., Phys. Rev. C42 (1990) R1791;
C.S. Wu et al., Phys. Rev. C45 (1992) 261;
J.A. Becker et al., Phys. Rev. C46 (1992) 889.
\bibitem{iac91} F. Iachello, Nucl. Phys. A522 (1991) 83c.
\bibitem{jan92} R.V.F. Janssens and T.L. Khoo, Ann. Rev. Nucl. Part. Sci. 41
(1991) 321.
\end{thebibliography}
\end{document}